\documentclass[prl,twocolumn,nofootinbib, preprintnumbers, superscriptaddress]{revtex4-1}

\usepackage{amsmath,amssymb,amscd,simplewick}
\usepackage{listings}
\usepackage{dsfont}
\usepackage{slashed}
\usepackage{color}
\usepackage{ulem}

\usepackage{graphicx}
\usepackage{epstopdf}
\usepackage{subfigure}
\usepackage{epsfig}

\usepackage{xcolor}
\usepackage[colorlinks=true,
            linkcolor=blue,
            urlcolor=blue,
            citecolor=green,          
            bookmarks=true,
            bookmarksnumbered=true,
            breaklinks=true,
            pdfpagemode=Fullscreen,
            pdfstartview=FitBH]{hyperref}

\hypersetup{pdfauthor = {Shao-Feng Ge},
	     pdftitle = {}, 
	     pdfsubject = {}, 
             pdfkeywords = {}, 
	     pdfcreator = {LaTeX with hyperref package},
	     pdfproducer = {dvips + ps2pdf} }

\definecolor{gesfpurple}{rgb}{0.47,0.19,0.42}

\definecolor{gesflanse}{rgb}{0.00,0.50,0.50}

\definecolor{gesfblue}{rgb}{0.08,0.42,0.76}

\definecolor{gesfred}{rgb}{1,0,0}

\definecolor{gesfwhite}{rgb}{1,1,1}

\definecolor{gesfblack}{rgb}{0,0,0}

\newcommand{\geqn}[1]{\hypersetup{linkcolor=blue}Eq.~(\ref{#1})\hypersetup{linkcolor=blue}}
\newcommand{\gfig}[1]{{\hypersetup{linkcolor=violet}Fig.~\ref{#1}\hypersetup{linkcolor=blue}}}

\graphicspath{{figs/}}

\begin{document}


\title{Diurnal Effect of Sub-GeV Dark Matter Boosted by Cosmic Rays}
\author{Shao-Feng Ge}
\email{gesf@sjtu.edu.cn}
\affiliation{Tsung-Dao Lee Institute, Shanghai Jiao Tong University, Shanghai 200240, China}
\affiliation{School of Physics and Astronomy, Shanghai Jiao Tong University, Key Laboratory for Particle Astrophysics and Cosmology (MOE) \& Shanghai Key Laboratory for Particle Physics and Cosmology, Shanghai 200240, China}
\author{Jianglai Liu}
\email{jianglai.liu@sjtu.edu.cn}
\affiliation{School of Physics and Astronomy, Shanghai Jiao Tong University, Key Laboratory for Particle Astrophysics and Cosmology (MOE) \& Shanghai Key Laboratory for Particle Physics and Cosmology, Shanghai 200240, China}
\affiliation{Tsung-Dao Lee Institute, Shanghai Jiao Tong University, Shanghai 200240, China}
\author{Qiang Yuan}
\email{yuanq@pmo.ac.cn}
\affiliation{Key Laboratory of Dark Matter and Space Astronomy, Purple Mountain Observatory, Chinese Academy of Sciences, Nanjing 210033, China}
\affiliation{School of Astronomy and Space Science, University of Science and Technology of China, Hefei 230026, China}
\affiliation{Center for High Energy Physics, Peking University, Beijing 100871, China}
\author{Ning Zhou}
\email{nzhou@sjtu.edu.cn}
\affiliation{School of Physics and Astronomy, Shanghai Jiao Tong University, Key Laboratory for Particle Astrophysics and Cosmology (MOE) \& Shanghai Key Laboratory for Particle Physics and Cosmology, Shanghai 200240, China}

\begin{abstract}
We point out a new type of diurnal effect for the
cosmic ray boosted dark matter (DM).
The DM-nucleon interactions not only allow
the direct detection of DM with nuclear recoils, 
but also allow cosmic rays to scatter with
and boost the nonrelativistic DM to higher energies.
If the DM-nuclei scattering cross sections are sufficiently
large, the DM flux is attenuated as it propagates 
through the Earth, leading to a strong diurnal modulation.
This diurnal modulation provides another
prominent signature for the direct detection of boosted sub-GeV DM, in addition to signals with higher recoil energy.
\end{abstract}

\maketitle 

{\it Introduction} --
Overwhelming evidence from astrophysical and cosmological
observations supports the existence of dark matter (DM)
\cite{Young:2016ala}, which is gravitationally interacting
but invisible via electromagnetic interactions. 
However, the physical nature of DM is poorly understood: 
the DM identity is unknown with a possible mass spans nearly 80 orders
of magnitude \cite{Gelmini:2015zpa}.
The DM direct detection \cite{Goodman:1984dc}
aims to verify the existence of 
DM particles and measure their interactions via the recoil of target nuclei or electrons,
which is believed to be the most direct 
way to unveil
the nature of DM particles \cite{Liu:2017drf,Schumann:2019eaa}.

Conventionally, direct detection experiments assume the existence of nonrelativistic DM confined in the Galaxy. 
The gravitational potential of the Galaxy
results in an upper limit on the DM velocity of
$v_\chi \lesssim 600~\mbox{km/s}$ above which DM can
escape \cite{Bozorgnia:2013pua,Zavala:2019gpq}.
Because of the energy threshold, which is typically
$\mathcal O(\mbox{keV})$, the sensitive mass
window of direct detection experiments can only 
extend down to $\mathcal O(1)$\,GeV via the conventional nuclear recoil channel.
In recent years, to enhance the sensitivity of detecting sub-GeV DM,
many approaches have been explored, including expanding
the nuclear recoil detection capability via a low threshold
bolometer \cite{Abdelhameed:2019hmk,Bolometer} as well as via the Bremsstrahlung \cite{Kouvaris:2016afs} and Migdal \cite{Ibe:2017yqa,Baxter:2019pnz,Essig:2019xkx,Akerib:2018hck,Armengaud:2019kfj,Liu:2019kzq,Aprile:2019jmx} effects, the direct detection of DM-electron recoils \cite{eRecoil,Emken:2019tni,Xenon-eRecoil,DarkSide-eRecoil,SuperCDMS-eRecoil,CCD}, and various novel detection proposals
\cite{SC,FD,SF,SH,Scintillator,DiracMaterial,Molecular,Diamond,nanoWire,Plasmon,magnon}.

Another interesting possibility has been recently pointed out: nonrelativistic DM 
can be boosted by cosmic rays (CRs) \cite{Cappiello:2018hsu,Bringmann:2018cvk} or
the solar reflection \cite{Kouvaris:2015nsa,An:2017ojc,Emken:2017hnp}.
As long as DM has finite interactions with matter, it is inevitable for the
nonrelativistic DM to be scattered and boosted by the energetic CRs. Although the flux of the CR-boosted DM (CRDM) is a tiny fraction compared to the nonrelativistic DM, it allows explorations of a certain parameter space of sub-GeV DM that was previously inaccessible \cite{Bringmann:2018cvk,Alvey:2019zaa,Dent:2019krz,Wang:2019jtk,Plestid:2020kdm} in direct detection, thus expanding the sensitive mass region.
The CRDM can also produce signals in large neutrino
experiments \cite{Ema:2018bih,Cappiello:2019qsw,Guo:2020drq}.

For sub-GeV DM, the DM-nucleon scattering cross section
with a contact interaction can be quite sizable, e.g.,
as large as
$10^{-31}\mbox{cm}^2$ (see \cite{Krnjaic:2019dzc}
and the references in \cite{Cappiello:2018hsu}),
in contrast to the light mediator case
\cite{Bondarenko:2019vrb}.
With this allowed interaction strength,
DM particles can
experience multiple scatterings and become
attenuated when traveling through the Earth
\cite{Starkman:1990nj, Mack:2007xj, Hooper:2018bfw, Emken:2018run}.
If the CRDM flux is anisotropic, a diurnal flux modulation at direct detection experiments is expected
\cite{diurnal,diurnal-directional}.
This is different from the conventional diurnal effect that is mainly for 
nonrelativistic DM.

{\it Sub-GeV Dark Matter Boosted by Cosmic Rays} --
The spatial and spectral distributions of the CRDM
flux depend on the DM and CR distributions in the 
Galaxy as well as the CRDM scattering processes.
Both the DM density and CR intensities vary with
their locations in the Galaxy, becoming more concentrated
toward the Galaxy center (GC). Therefore, CRs are much more
likely to scatter with and boost the DM in the inner Galaxy
region. Even for isotropic scattering, the CRDM flux 
is highly anisotropic over the sky.

Although the CRDM scattering also affects the CRs, the effect 
is important only for a very large scattering cross section
($\sigma_{\chi p}>10^{-27}$~cm$^2$) \cite{Cappiello:2018hsu}.
For simplicity, we assume that the CR distribution is unaffected. The CRDM emissivity, which describes 
its spatial and spectrum distributions, is given by 
\cite{Bringmann:2018cvk}
\begin{eqnarray}
  \zeta_\chi({\bf r},T_{\chi})
& = &
  \frac{\rho_{\chi}(|{\bf r}|)}{m_{\chi}}
  \sum_{i=p,{\rm He}}
  \int_{T_i^{\rm min}}^{\infty}
  {\rm d}T_i\frac{n_{{\scriptsize \rm CR},i}({\bf r},T_i)}{T_{\chi}^{\rm max}(T_i)}
\nonumber\\
&&
\hspace{27mm}
\times
  v_i \sigma_{\chi i}G_i^2(Q^2),
\label{eq:emissivity}
\end{eqnarray}
where $T_i$ and $T_\chi$ are the kinetic energies
of the CR species $i$ and the boosted DM
with mass $m_\chi$,
$T_i^{\rm min}$ is the minimum CR energy 
required to boost the DM kinetic energy to $T_{\chi}$,
and $T_{\chi}^{\rm max}$ is the maximum DM kinetic
energy given $T_i$
\cite{Bringmann:2018cvk}.
There are three main ingredients in \geqn{eq:emissivity}:
the DM density $\rho_\chi(|{\bf r}|)$ at
location ${\bf r}$, the CR density
$n_{{\scriptsize \rm CR},i}$ times its velocity
$v_i$, and the
scattering cross section $\sigma_{\chi i}$.
The form factor $G_i(Q^2) \equiv 
1/(1+Q^2/\Lambda_i^2)^2$ \cite{Perdrisat:2006hj}
is a function of the momentum transfer $Q$
with $\Lambda_p\approx 770$~MeV 
and $\Lambda_{\rm He}\approx 410$~MeV \cite{Angeli:2004kvy} for
proton and helium, respectively.

\begin{figure}[t]
\centering
\includegraphics[width=0.48\textwidth,height=5cm]{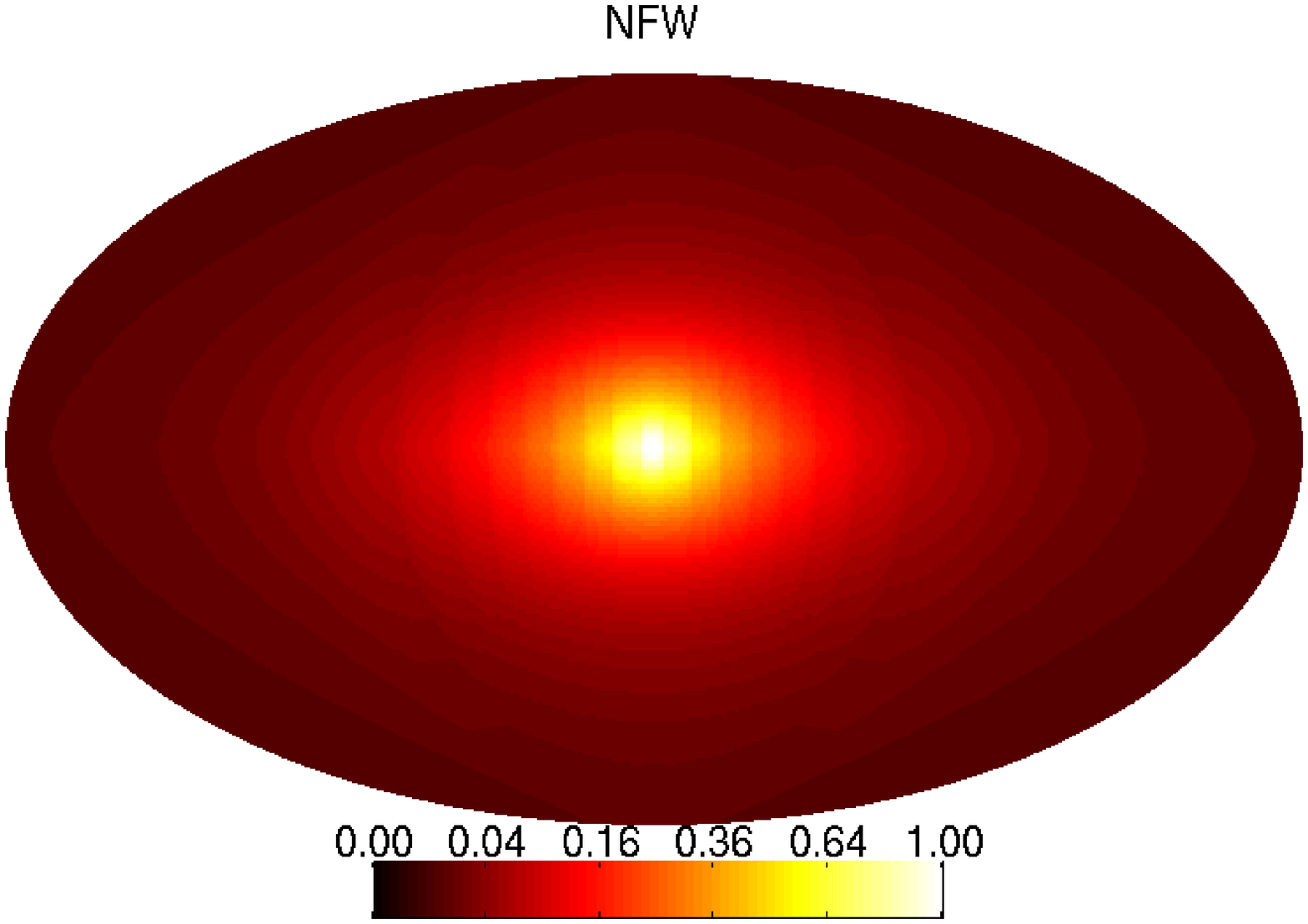}
\\[3mm]
\includegraphics[width=0.48\textwidth,height=5cm]{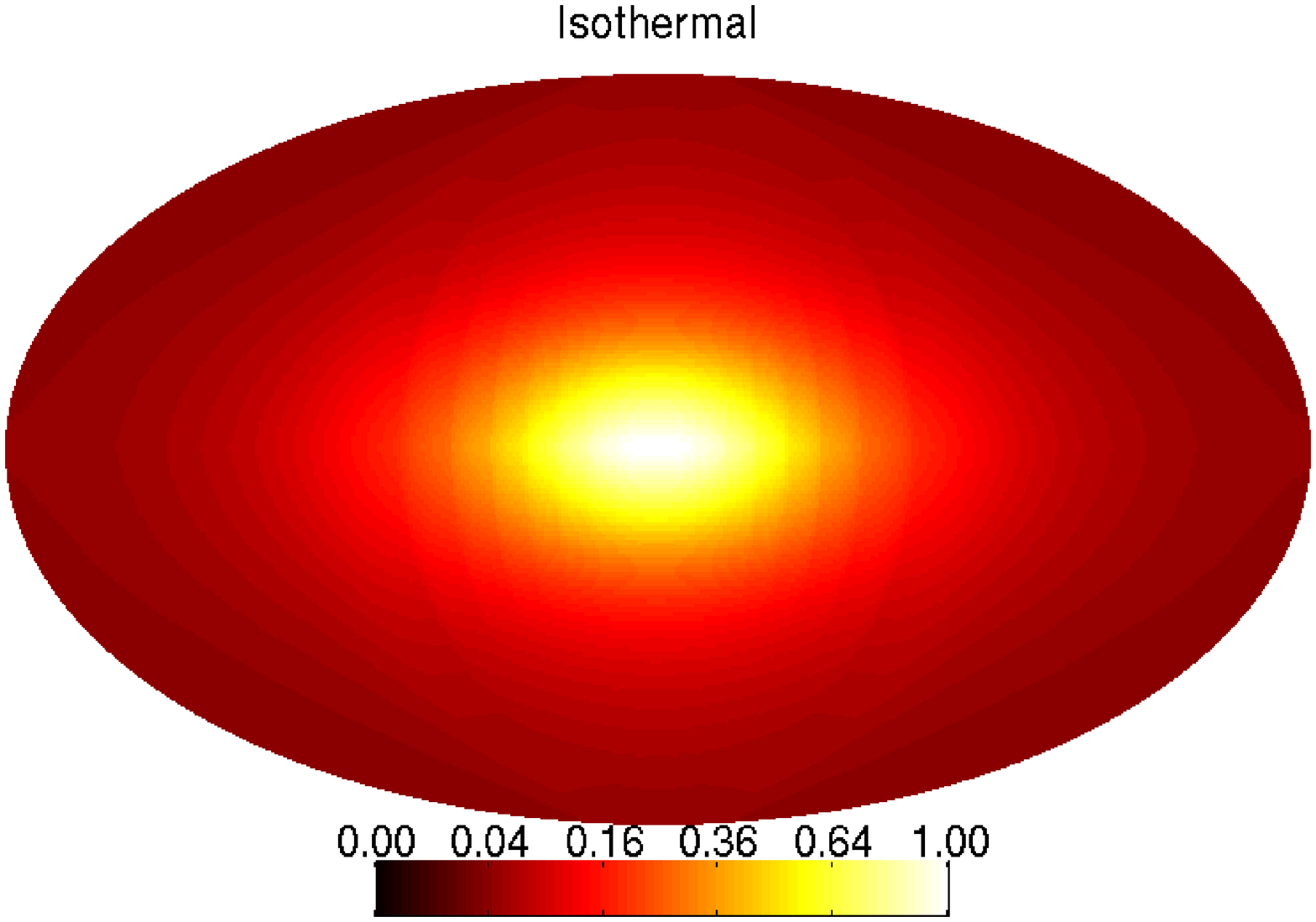}
\caption{
Relative sky maps of CRDM fluxes in the Galactic coordinates with amplitude in the GC direction set to unity.
The upper and lower panels are for the NFW and Isothermal DM density profiles, respectively.}
\label{fig:skymap}
\end{figure}

For the DM density $\rho_\chi(|{\bf r}|)$, 
we adopt the Navarro-Frenk-White (NFW) \cite{Navarro:1996gj}
profile, 
$\rho_{\chi}^{\rm nfw}(r)=\rho_s/[(r/r_s)(1+r/r_s)^2]$
with $r_s=20$~kpc and $\rho_s=0.35$~GeV~cm$^{-3}$,
 as the benchmark DM mass distribution.
For comparison, a cored isothermal distribution,
$\rho_{\chi}^{\rm iso}(r)=\rho_s/[1+(r/r_s)^2]$
with $r_s=5$~kpc and $\rho_s=1.56$~GeV~cm$^{-3}$, is also studied.
These parameters correspond to a local DM density of
$0.4$~GeV~cm$^{-3}$ in our Solar System \cite{Catena:2009mf} for both profiles. 
The difference between the two profiles and more details
are given in the Supplemental Material \cite{Supplementary}.
The amplitudes of the diurnal modulation vary
by only around $7\%$ for different density profiles.

For the CR contribution in \geqn{eq:emissivity}, 
we employ the GALPROP \cite{Strong:1998pw} 
code (version 54) to simulate its distribution.
In this Letter, we only consider the dominating proton
and helium species of CRs, and leave the rest, in particular
electrons and positrons, for future discussions.
For the detailed CR model parameters and the resulting CR spatial
distribution, please see the 
Supplemental Material \cite{Supplementary}.

The DM-nucleus interaction is the least known part in \geqn{eq:emissivity}.
For simplicity, we assume that 
the DM-nucleus cross section $\sigma_{\chi A}$ has a
coherent enhancement,
\begin{equation}
  \sigma_{\chi A}
=
  \sigma_{\chi p} A^2
\left[ \frac{m_A(m_{\chi}+m_p)}{m_p(m_{\chi}+m_A)} \right]^2,
\end{equation}
where $\sigma_{\chi n}=\sigma_{\chi p}$ is the constant
DM-nucleon cross section, while $m_p$ and $m_A$
are the proton and nuclear masses for the CR.
For $m_\chi \ll m_p, m_A$, the enhancement mainly
comes from the $A^2$ factor. Extra enhancement may come
from $(m_\chi + m_p)^2 / m^2_p$ when $m_\chi$ goes beyond
$m_p$. 
The dipole hadronic form factor $G_i(Q^2)$ in
\geqn{eq:emissivity} suppresses the interaction at
large momentum transfer $Q$.

\begin{figure}[t]
\centering
\includegraphics[width=0.45\textwidth]{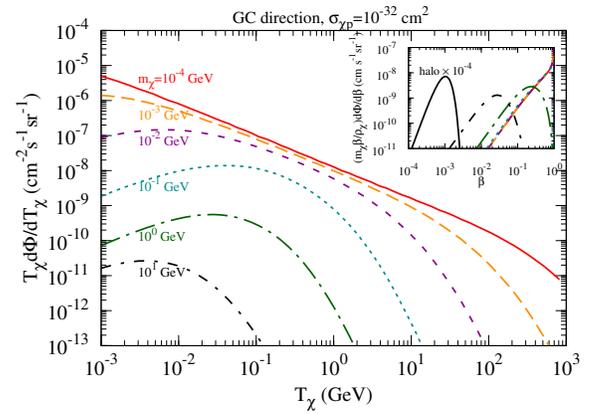}
\caption{The CRDM energy spectra at the GC direction
for DM masses $10^{-4}$, $10^{-3}$, $10^{-2}$, 0.1, 
1.0, and 10 GeV from top to bottom. The scattering cross 
section $\sigma_{\chi p}$ is assumed to be $10^{-32}$~cm$^2$.
The inset is the distribution of DM velocities, 
$\beta=v/c$, compared to the Maxwellian distribution of 
the Standard DM Halo. For a clear comparison,
we rescale the Standard DM Halo curve by
$10^{-4}$ (labeled as halo $\times 10^{-4}$
in the inset) so that all curves have a similar
height.}
\label{fig:spec_crdm_gc}
\end{figure}

The CRDM flux arriving at the Earth along a given direction
$\hat{\bf n}$ is a line-of-sight integral of all
contributions along the way,
\begin{equation}
    \frac{{\rm d}\Phi}{{\rm d}T_{\chi}}({\bf \hat{n}},T_{\chi})=\frac{1}{4\pi}
    \int \zeta_\chi({\bf r},T_{\chi})\,{\rm d}l.
\end{equation}
\gfig{fig:skymap} shows the relative all-sky maps of the CRDM fluxes in the Galactic coordinate, a spherical coordinate with the Sun as its center, the latitude measuring the angle above/below the galactic plane, and the longitude measuring the azimuth angle from the GC. The peak value at the GC is set to 1. The top (bottom) panel presents 
the NFW (isothermal) profile.
The CRDM fluxes are clearly anisotropic, with the maximum 
(the GC direction) and the minimum
differing by about two orders of magnitude. 
To match the grid 
resolution of GALPROP, we set the NFW density within 
0.5 kpc of the GC to $\rho(0.5~{\rm kpc})$.
This approximation has a negligible effect on
the diurnal modulation, 
as shown in the
Supplemental Material \cite{Supplementary}.

\gfig{fig:spec_crdm_gc} shows the CRDM spectra
from the GC direction for different DM masses.
The number density $\rho_\chi/m_\chi$ in \geqn{eq:emissivity}
accounts for the decrease of CRDM flux for larger DM masses.
On the other hand, on average the maximum boost occurs when
$m_\chi$ approaches the mass of the incident proton or helium,
manifesting in the change of spectrum shape for different
energies. 
At the high energy end, the spectra are 
suppressed by the form factor $G_i(Q^2)$ with
$Q^2 = 2 m_\chi T_\chi$.
We also show the nonrelativistic DM velocity
distribution predicted by the Standard DM
Halo model (labeled as halo $\times 10^{-4}$) in
\gfig{fig:spec_crdm_gc} for comparison.

We find that the CRDM spectra depend very weakly on directions, 
mainly due to the similar CR spectral shapes throughout the Galaxy.
For simplicity, in the following discussion we will separate the
energy and angular distributions of the CRDM fluxes.

{\it Earth Attenuation} --
With a large enough scattering cross section, the DM can frequently
scatter with matter when traveling through the Earth 
\cite{Starkman:1990nj, Mack:2007xj, Hooper:2018bfw, Emken:2018run,Emken:2019tni}, 
transferring its kinetic energy to matter nuclei.
Although the decelerated DM particle may still reach the
detector, the DM energy spectrum is shifted lower,
leading to fewer events above the detector energy threshold.
For simplicity, we use the average nucleon numbers, $\bar A_m = 24$ 
in the Earth mantle and $\bar A_c = 54$ in the Earth core, to approximate the 
matter compositions \cite{McDonough:2008zz}. As a concrete example, for 
$\sigma_{\chi p} = 10^{-32}~\mbox{cm}^2$, the mean free path,
$L_{\rm free} \equiv m_N / (\rho_N \sigma_{\chi A})$, is
around $2.7/17$~km in the Earth core/mantle 
omitting the form factor effects.
Similar attenuation happens in the atmosphere, 
but due to the 3 orders lower density, the effect 
is only visible at much larger cross sections.

\begin{figure}[t!]
\centering
\includegraphics[width=0.45\textwidth,height=55mm]{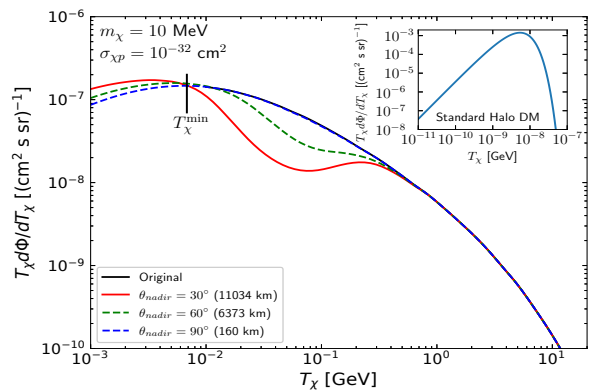}
\caption{The attenuated CRDM spectra for the nadir angles $\theta_{\rm nadir} = 30^\circ$ (red), $60^\circ$ (green), and $90^\circ$ (blue) 
with $\sigma_{\chi p} = 10^{-32}~\mbox{cm}^2$, $m_\chi = 10\,\mbox{MeV}$, and the detector 
at a depth of 2\,km. For comparison, we also
show the original Standard Halo DM flux 
distribution in the inset.}
\label{fig:single}
\end{figure}

The differential CRDM flux $d \Phi({\bf \hat{n}}, l, T_\chi) / d \ln T_\chi$,
at the distance $l$ through the Earth, is a combination of
the loss of DM particles to an energy lower than $T_\chi$
and the gain from a higher energy $T'_\chi$ to $T_\chi$. 
For an incoming DM particle with a higher energy $T'_\chi$, 
the nuclear recoil energy $T_r$ is evenly distributed in the range
$0 \leq T_r \leq T'_\chi (T'_\chi + 2 m_\chi) /
(T'_\chi + m_\mu) \equiv T^{\rm max}_r(T'_\chi)$
with reduced mass $m_\mu \equiv (m_N + m_\chi)^2 / 2 m_N$.
Because of energy conservation, $T_\chi$ is also evenly distributed:
$T'_\chi (m_\mu - 2 m_\chi) / (T'_\chi + m_\mu) \leq T_\chi
\leq T'_\chi$. For a given $T_\chi$, the DM particles with energy
$T'_\chi$ in the range $T_\chi \leq T'_\chi \leq m_\mu T_\chi /
(m_\mu - 2 m_\chi - T_\chi)$ increases the flux at $T_\chi$.
The CRDM flux evolution
contains two contributions~\cite{Ema:2018bih}:
\begin{eqnarray}
&&
  \frac \partial {\partial l}
  \frac {d \Phi(l, T_\chi)}{d \ln T_\chi}
=
  \frac {\rho_N(l)}{m_N} \sigma_{\chi N}
\left[
- \frac {d \Phi(l, T_\chi)}{d \ln T_\chi}
  w_{\rm FF}(T_\chi)
\right.
\nonumber
\\
&&
\hspace{8mm} +
\left.
  \int
  \frac {d \Phi \left(l, T'_\chi \right)}
        {d \ln T'_\chi}
  \frac {T_\chi (T'_\chi + m^N_\mu)}
		{T'_\chi (T'_\chi + 2 m_\chi)}
  G^2_N(Q^2) d \ln T'_\chi
\right].
\quad
\label{eq:IDE}
\end{eqnarray}
The weight factor is defined as,
$w_{\rm FF} \equiv \int G^2_N(Q^2) d Q^2 / Q^2_{\rm max}$,
and the factor $T_\chi / T^{\rm max}_r$
in the second term comes from the
differential cross section $d \sigma
= \sigma d T_r / T^{\rm max}_r
= \sigma d \ln T_\chi (T_\chi / T^{\rm max}_r)$.
The attenuated DM flux can be obtained by integrating
\geqn{eq:IDE} step by step over the traversed distance.
\gfig{fig:single} shows the attenuated CRDM fluxes with
different nadir angles to the underground detector. To be
realistic, we consider a detector $2~\mbox{km}$ underground.
Then for $\theta_{\rm nadir} = 90^\circ$, DM needs to travel
$160~\mbox{km}$ before reaching the detector, corresponding
to 9 mean free paths in the mantle.
The CRDM
flux at medium energy is largely reduced first and then 
goes back up at high energy. The limited attenuation at
high energy is due to the highly suppressed 
weight factor $w_{\rm FF}(T_\chi)$ in \geqn{eq:IDE}.
Consequently,
the CRDM is much more energetic
than the nonrelativistic DM (see the inset of \gfig{fig:single}) and can 
produce recoil events with much higher energy. This makes direct detection experiments sensitive to sub-GeV DMs.

{\it Boosted Diurnal Effect} --
The two anisotropies from the Earth and the Galaxy lead to the diurnal effect. First, the path 
lengths that DM particles traverse are anisotropic since the 
underground lab is close to the Earth surface and its depth is 
typically much smaller than the Earth radius. 
Second, the CRDM flux is strongly peaked toward the GC due to 
both the DM and the CR distributions. 
The CRDM flux is thus significantly attenuated by the Earth when
the GC and the detector are on opposite sides of the Earth
but much less affected if they are on the same side. 
To avoid confusion with the usual 
diurnal effect for nonrelativistic DM
\cite{diurnal,diurnal-directional}, 
we call this the ``{\it boosted diurnal effect}''.

\begin{figure}[t!]
\centering
\includegraphics[width=0.45\textwidth,height=57mm]{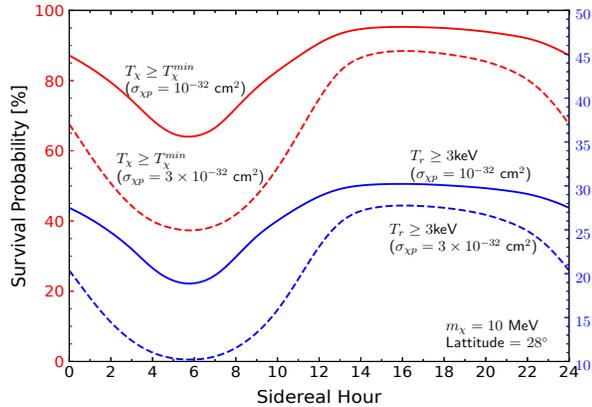}
\caption{The survival probability of CRDM arriving at an
underground lab at latitude $28^\circ$N and a depth of
$2\,\mbox{km}$ vs the sidereal hour,
relative to the number of DM particles arriving at the Earth for two different cross sections
$\sigma_{\chi p} = 1~(3) \times 10^{-32}$~cm$^2$. The red
curves correspond to the total CRDM arriving at the detector
with $T_\chi \geq T^{\rm min}_\chi$, and the blue curves are
those above the detector threshold ($T_r > 3\,\mbox{keV}$
for a liquid xenon detector).}
\label{fig:lab}
\end{figure}

\gfig{fig:lab} shows the diurnal modulation of the CRDM at a direct
detection experiment located at a latitude of $28^\circ$N
 (approximate location of the China Jinping 
Underground Laboratory) and a depth
of 2~km underground. Within one sidereal day, the underground lab 
rotates around the Earth axis and its position is parameterized by the 
sidereal hour in the range between $[0, 24]$ hours. We define a survival
probability as the ratio between the attenuated CRDM flux in the underground lab and the one arriving the Earth.
At a cross section of $1\times 10^{-32}$~cm$^2$, we observe significant
``{\it boosted diurnal modulation}'' with the survival probability varying 
in the range of $64\% \sim 95\%$. For comparison, we also show
the curves for a cross section of $3\times 10^{-32}$~cm$^2$ 
where a larger modulation can be observed.
Given the DM energy $T_\chi$, the nuclear recoil
has a wide distribution,
$0 \leq T_r \leq T^{\rm max}_r(T_\chi)$,
and hence only a fraction,
$1 - T_{th} / T^{\rm max}_r(T_\chi)$, can pass
the detection threshold, leading to a reduction
from the red curve to the blue one in
\gfig{fig:lab}.

Instead of performing numerical integration of \geqn{eq:IDE}, the curves in \gfig{fig:lab} are obtained by Monte Carlo simulations.
Since the spectrum of the CRDM is almost independent of its direction, 
it is a good approximation to first sample the direction of the 
incoming DM particles according to the sky map
in \gfig{fig:skymap} and then sample the boosted DM kinetic energy $T_\chi$
according to the spectrum in \gfig{fig:spec_crdm_gc}. The incident DM 
particle would then experience multiple scatterings
when crossing the Earth.
For each interaction step, we first sample the distance that the DM 
particle travels before the next scattering based on the mean free path and then sample the reduced kinetic energy. 
The simulation stops when the DM particle reaches the underground detector or drops below the
detection threshold.

Imposing the detection threshold on the
nuclear recoil energy,
$T_r \geq 3\,\mbox{keV}$ for a liquid xenon detector
\cite{Cui:2017nnn}, would reduce the event
rate but still keep the modulation behavior
as illustrated in \gfig{fig:lab}.
This is because the diurnal modulation mainly
comes from the high recoil part as illustrated
in \gfig{fig:single}.
For two years of data at a benchmark liquid
xenon detector PandaX-4T (5.6 ton$\times$year
exposure) \cite{Zhang:2018xdp}, on average
8.1 (55) events are expected 
for $\sigma_{\chi p} = 1~(3) \times 10^{-32}\,\mbox{cm}^2$
and $m_\chi = 10\,\mbox{MeV}$, which is quite significant 
compared to the background level~\cite{Wang:2019opt}.
For the same detector, the event rate and hence the sensitivity is roughly
independent of the DM mass for $m_\chi \lesssim 0.1\,\mbox{GeV}$.
In addition to a quadratic scaling with the cross section, one from the CRDM production and the other from its detection, the event rate is suppressed once the attenuation from the Earth becomes dominating for a sufficiently large cross section ($\sim10^{-28}$ cm$^2$) \cite{Bringmann:2018cvk}.
The cross section region that this technique can probe spans 
roughly 4 orders of magnitude.

Another factor is the scattering angle,
which leads to deflection \cite{Emken:2019tni}.
For the relativistic CRDM with typical 1\,GeV
kinetic energy, mass $m_\chi = 10$\,MeV, and
typical momentum transfer
$Q \approx \Lambda \approx 200$\,MeV \cite{Angeli:2004kvy},
the scattering angle is $3^\circ \sim 5^\circ$.
Although not completely negligible, the scattering
angle does not affect the diurnal modulation
effect due to the following arguments. For the peak region of \gfig{fig:lab},
the DM from the GC only needs to penetrate
$\mathcal O(1)\,\mbox{km}$. With a mean free path 
of around 17\,km, most CRDMs experience only 
one scattering at most. Therefore, the peak region 
would not be affected significantly.
Multiple scatterings will further suppress the valley region of the 
curve and therefore enhance the modulation effect.

\begin{figure}[t]
\centering
\includegraphics[width=0.45\textwidth,height=58mm]{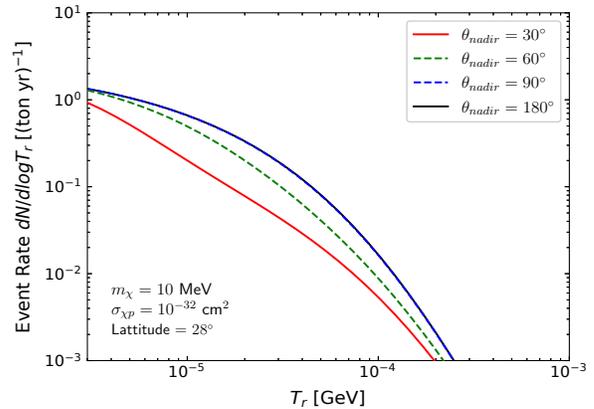}
\caption{The nuclear recoil spectrum, including
the 3\,keV detector threshold,
for a xenon detector with 1 ton$\cdot$year
exposure. To illustrate the attenuation
effect, each curve corresponds to the
integrated DM flux at a given nadir angle
$\theta_{\rm nadir}$.}
\label{fig:TrEff}
\end{figure}

The recoil energy spectra for incident CRDMs along different nadir angles in a liquid xenon detector
are shown in \gfig{fig:TrEff}. 
Since the recoil energy can reach $\mathcal O(1\,\mbox{MeV})$, 
observing a high energy recoil event is a
smoking gun for the CRDM, especially when the detector and the GC are 
on the same side of the Earth.
However, these energetic recoils may excite target
isotopes and therefore may no longer be simple nuclear
recoils. The signal identification strategy for
such events needs more experimental study.
Statistically, the boosted diurnal 
modulation can help to identify such high energy recoil signal and 
suppress the background which is expected to be constant over time.
A more detailed analysis with real data will
appear in a future work.

{\it Conclusion} --
The CRDM provides a possibility for the conventional
DM direct detection experiments to extend their 
sensitive window to the sub-GeV mass range via the 
detection of boosted DM events that produce a higher 
energy recoil above threshold. If the DM-nucleon cross 
section is sufficiently large, the CRDM is significantly 
attenuated when traveling through the Earth. 
Because of the anisotropies of the CRDM flux and the 
Earth attenuation, the event rate and energy spectrum
exhibit a characteristic diurnal modulation, which is 
a powerful signature to suppress
background and enhance sensitivities to sub-GeV DM.
Future work can use the electron component in the CR 
and extend this exploration to DM-electron interactions.
In addition, future directional detection experiments may
directly image the anisotropic sky map of the CRDM.
The modulation discussed in this Letter may also
apply to the boosted DM scenario \cite{Agashe:2014yua,Berger:2014sqa,Cherry:2015oca}.

\section*{Acknowledgements}
SFG is sponsored by the Double First Class start-up
fund (WF220442604) provided by Tsung-Dao Lee Institute \&
Shanghai Jiao Tong University, the Shanghai Pujiang
Program (20PJ1407800), and the National Natural Science
Foundation of China (No. 12090064).
QY is supported by the National Natural Science Foundation of China 
(Nos. 11722328 and 11851305), the 100 Talents Program of the Chinese 
Academy of Sciences, and the Program for Innovative Talents and 
Entrepreneur in Jiangsu.
JL and NZ are supported by the National Natural Science Foundation of 
China (Nos. 11525522, 11775141, and 12090061), a grant from the Ministry of Science 
and Technology of China (No. 2016YFA0400301), and a grant from Office 
of Science and Technology, Shanghai Municipal Government (No.
18JC1410200) and the Hongwen Foundation in Hong Kong. JL also 
acknowledges support from the Tencent Foundation. 
This work is also supported in part by the
Chinese Academy of Sciences Center for Excellence in Particle Physics (CCEPP).
The authors are grateful to Dr. Xiang-Yi Cui for checking the calculations.

\newpage

\section*{Supplemental Material}
\label{sec:supplementary}

\subsection{Dark matter density profile}

\begin{figure}[h]
\centering
\includegraphics[width=0.48\textwidth]{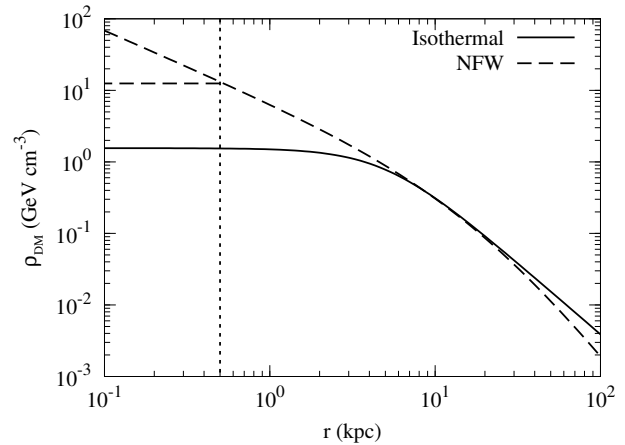}
\caption{The NFW and Isothermal density profiles of DM in the Milky
Way. The vertical dotted line shows the cut radius of 0.5 kpc 
within which the DM density is taken to be constant.}
\label{fig:rhoDM}
\end{figure}

Two kinds of density profiles, a cuspy NFW one and a cored 
isothermal one are employed in this work. Fig.~\ref{fig:rhoDM} shows 
the density distributions of both profiles. To match with the grid 
precision of the GALPROP propagation model, we employ a cut radius 
for the calculation of the NFW density profile, within which the 
density is taken as a constant. We have tested that for $r_c=0.2$, 
0.5, and 1.0 kpc, the final results differ very little.

\subsection{Cosmic ray modeling}

Charged CRs propagate diffusively in the random magnetic
field of the Milky Way. Besides the diffusion effect, particles
may get re-accelerated by randomly moving magnetic turbulence, 
advected from the Galactic disk to the halo, and interact with 
the medium. It has been shown that the diffusion plus reacceleration 
propagation framework can best match the current data \cite{Yuan:2018lmc}. 
The propagation parameters are obtained by fitting the most 
recent AMS-02 measurements of the secondary and primary nuclei 
\cite{Aguilar:2017hno,Aguilar:2018njt}. 
The propagation parameters adopted are: the diffusion 
coefficient $D(R)=\beta^{\eta}D_0(R/{4~{\rm GV}})^{\delta}$ 
with $D_0=7.13\times10^{28}$~cm$^2$~s$^{-1}$, $\delta=0.353$,
$\eta=0.0$, the half-height of the propagation cylinder 
$z_h=5.4$~kpc, and the Alfven velocity $v_A=35.4$~km~s$^{-1}$ 
that characterizes the reacceleration effect. 

The spatial distribution of CR sources is assumed to follow
that of supernova remnants \cite{Case:1998qg}. For the source
spectra of CRs, we employ a spline-interpolation approach as
in Ref.~\cite{Yuan:2018lmc}. The spectral parameters are derived 
through fitting to the proton and helium data which are measured 
in a wide energy range from outside of the Solar System by 
Voyager-1 \cite{Cummings:2016pdr}, and at the top-of-atmosphere 
by AMS-02 \cite{Aguilar:2015ooa,Aguilar:2015ctt}, 
CREAM-III \cite{Yoon:2017qjx}, and DAMPE \cite{An:2019wcw}.
To connect the local interstellar spectra of CRs with the 
measured ones around the Earth, a force-field solar 
modulation model is applied \cite{Gleeson:1968zza}.

\begin{figure}[t]
\centering
\includegraphics[width=0.48\textwidth]{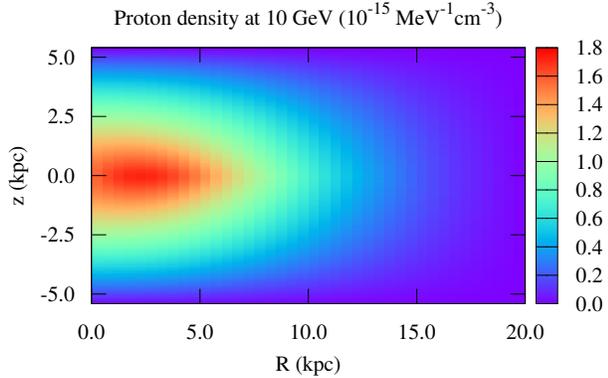}
\caption{Density distribution of 10 GeV CR protons in the Milky Way
as a function of radius $R$ and height $z$.}
\label{fig:proton_den}
\end{figure}

Fig.~\ref{fig:proton_den} shows the distribution of protons
at energies of 10 GeV, calculated with the GALPROP tool.
It shows that the CR density is not uniform, and higher
in the inner Galaxy. The product of the CR density and the
DM density thus gives a strong anisotropy of the expected
CRDM fluxes, as shown in Fig.~\ref{fig:skymap}.

\begin{center}
{\bf Effect on Diurnal Effect}
\end{center}

\begin{figure}[h]
\centering
\includegraphics[width=0.48\textwidth]{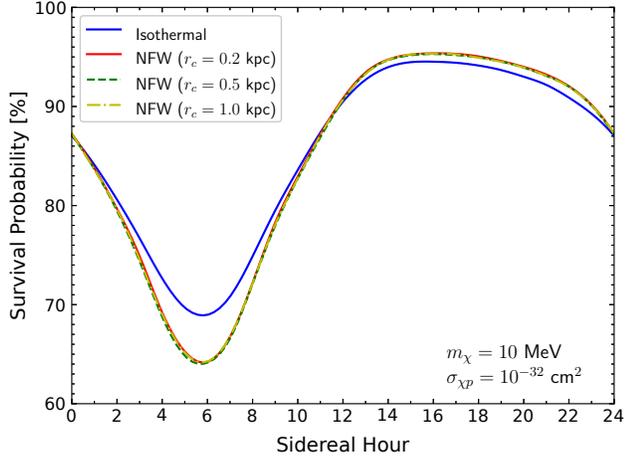}
\caption{The effect of DM profiles on the diurnal
modulation.} 
\label{fig:labFluxProfiles}
\end{figure}

Fig.~\ref{fig:labFluxProfiles} compares the effects of different
density profiles and smooth scales of the DM distribution on the
diurnal modulation effect.
We can see that for different DM profiles,
the basic features of diurnal modulation is
not affected with only slight change. The largest difference
between Isothermal and NFW profiles is around $7\%$
at the modulation valley.

\end{document}